# A simple method to downscale daily wind statistics to hourly wind data

Zhongling Guo

(College of Resource and Environmental Sciences/Hebei Key Laboratory of
Environmental Change and Ecological Construction, Hebei Normal University,
Shijiazhuang 050024)

**ABSTRACT:** Wind is the principal driver in the wind erosion models. The hourly wind speed data were generally required for precisely wind erosion modeling. In this study, a simple method to generate hourly wind speed data from daily wind statistics (daily average and maximum wind speeds together or daily average wind speed only) was established.

A typical windy location with 3285 days (9 years) measured hourly wind speed data were used to validate the downscaling method. The results showed that the overall agreement between observed and simulated cumulative wind speed probability distributions appears excellent, especially for the wind speeds greater than 5 m s$^{-1}$ range (erosive wind speed). The results further revealed that the values of daily average erosive wind power density (AWPD) calculated from generated wind speeds fit the counterparts computed from measured wind speeds well with high models' efficiency (Nash-Sutcliffe coefficient). So that the hourly wind speed data can be predicted from daily average and maximum wind speed data or only daily average wind speed data, respectively. Further studies are needed to examine the findings for inter-site wind data.

## Introduction

Wind erosion play an important role in shaping the Earth's surface. The need to estimate soil erosion by wind yields many wind erosion models (Zobeck et al, 2003; Webb and McGowan, 2009). Wind data with high temporal resolution is fundamentally important in precisely wind erosion modeling. For example, the Wind Erosion Prediction System (WEPS) and Revised Wind Erosion Equation (RWEQ) generally require hourly wind series (Hagen, 1996; Fryrear et al., 1998). Van Donk et al. (2008) demonstrated that four-wind data per day (measured at LT 0200, 0800, 1600, 2000) are suitable for use in WEPS while Guo et al. (2013 ) revealed that the same type of wind data can be used to evaluate wind erosion potential in RWEQ. However, hourly wind data or four-wind data per day are not always available for some locations. Meteorological observations with daily wind statistics may only be available for some sites.

One naturally wonder whether one can directly use daily average wind data to estimate period wind erosion for WEPS and RWEQ. Namikas et al. (2003) have shown that the intervals of wind data affect shear velocity estimates and possibly further influence predicting aeolian transport sediment flux. Larsén and Mann (2006) have shown that the wind speed with long averaging time can remove substantial extreme gusts. Some of the gusts may be above threshold and crucial to evaluate wind erosivity. Guo et al. (2012) have shown that the periods of wind speed can significantly under-estimate values of wind erosivity, and may further under-predict soil loss. These studies indicate that the type of wind data have significant impacts on wind erosion prediction. Thus, we cannot directly use wind data averaged over a day in WEPS and RWEQ.

One also might ask whether one can convert longer-period (such as a day) wind speed data to hourly wind speed data. For accuracy estimates of wind erosivity, the wind speeds greater than threshold (erosive wind speeds) are critical. In practice, many studies have been performed to make converting between different wind data types. A gust factor (or Durst Curve) is generally used to predict the extreme gust for a period (Durst, 1960; Deacon, 1965). Dynamical and statistical methods are used to downscale monthly or daily wind data for global climate simulations (Sailor et al., 2008; Pavlik et al., 2012). But these conversion method mentioned above are unable to evaluate some gusts, which may be above threshold and critical to wind erosion modeling. Several stochastic or deterministic wind generation methods have been developed for describing diurnal pattern of wind speed (Peterson and Parton, 1983; Skidmore and Tatarko, 1990; Ephrath et al., 1996; Fryrear et al., 1998; Donatelli et al., 2009). Yet these simulation equations need historical wind statistics, which are obtained from hourly or more detail wind data.

Where only daily wind statistics (such as daily average wind data) are available, the challenge for application of the wind erosion models requiring hourly wind data, such as WEPS and RWEQ, to various sites still remains. The purpose of this study is to develop a method of downscaling the daily wind statistics to hourly wind series for wind erosion modeling.

**Materials and Method**

For some sites, only some daily wind statistics are saved. Here we discussed two general case: (1) the daily average, maximum wind speed are available, how to use the two daily wind statistics to predict the magnitude of hourly wind within a day. The diurnal variation for the magnitude of hourly wind speeds can be calculated by:

$$W_n = W_{ave} + \frac{1}{\pi} W_{max} COS(\frac{n \cdot \pi}{12}) \qquad (1)$$

where $W_n$ is wind speed at hour of *n*, $W_{ave}$ is the daily average wind speed, $W_{max}$ is the daily maximum wind speed. 24 wind speeds were reproduced for a day using Equation 1 (*n* varies from 0 to 23); (2) In some cases, only daily average wind speed data are available, the hourly wind speeds variation can be adjusted as:

$$W_n = W_{ave} + \frac{1}{2} W_{ave} COS(\frac{n \cdot \pi}{12}) \qquad (2)$$

The other steps are as same as that mentioned above. Accordingly, we can also obtain 24 wind speeds from Equation 2. Occasionally, the value of $W_n$ generated by Equation 1 or 2 may be less than zero, in this case, the $W_n$ was assumed as 0 (calm).

The measured wind data were used to validate the downscaling methods. The hourly wind data are obtained from the Plant Stress & Water Conservation Meteorological Tower, USDA-ARS Cropping Systems Research Laboratory in Lubbock County, Texas (Figure 1). The sampling site lies on the level plains of the Llano Estacado, a region known for its windy conditions and associated wind erosion problems (Stout, 2010). Wind speed was measured at a height of 2 m using a propeller-type anemometer. The selected sampling period extended over a period of 9 years from January 1, 2001 to December 31, 2009. During this time, only 2 day (Feb. 29th, 2004 and Feb. 29th, 2008) of wind data were lost. The daily wind statistics (daily average and maximum wind speeds) dataset is extracted from the hourly average wind speed data.

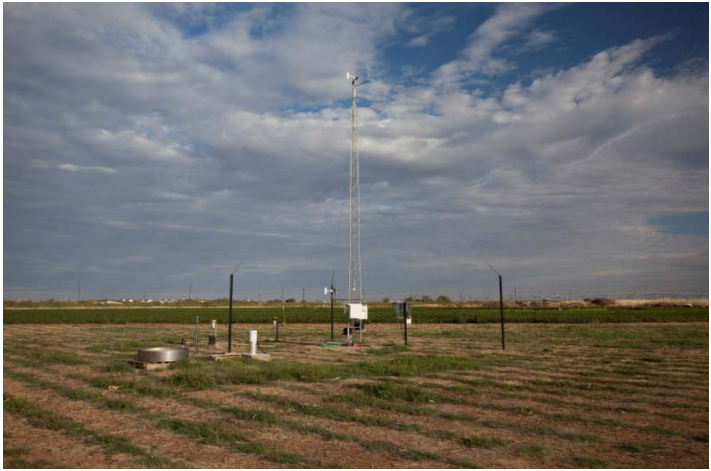

Figure 1. The Plant Stress & Water Conservation Meteorological Tower placed on the experimental field of the USDA-ARS Cropping Systems Research Laboratory in Lubbock County, Texas.

Compared with wind speed, the wind erosivity describing the potential of wind to generate sediment transport is more closely related to soil loss rate by wind (Shao, 2008). In this paper, a widely used wind erosivity formula defined as the erosive wind power density (WPD) (Lettau and Lettau, 1978; Greeley and Iverson, 1985; Hagen, 1996; Van Donk et al., 2008) was chosen to further evaluate how well wind is generated by Equation 1 and 2. The WPD is calculated by (Van Donk et al., 2008):

$$WPD = \frac{1}{2}\rho(U - U_t)U^2 \qquad (3)$$

where WPD is erosive wind power density (W m$^{-2}$), U is wind speed (m s$^{-1}$) and $U_t$ is threshold wind speed (m s$^{-1}$). When wind speed is below the threshold wind speed, WPD is zero and there is no sediment transport. The average WPD is computed based on wind statistics (Van Donk, et al., 2008). For each estimation period, the average WPD can be calculated by:

$$AWPD = \int_{U_t}^{U_{max}} P(U)\, WPD\, dU \qquad (4)$$

where AWPD is the average WPD (W m$^{-2}$), $U_{max}$ is the maximum wind speed (m s$^{-1}$), and P(U) is the wind speed probability for the estimation period. Here the threshold wind speed ($U_t$) and air density ($\rho$) were assumed as fixed values of 5 m s$^{-1}$ and 1.293 kg m$^{-3}$, respectively.

The Nash-Sutcliffe coefficient (Nash and Sutcliffe, 1970) was used to access the two equations' efficiency for daily AWPD estimation.

$$NSC = 1 - \frac{\sum_{i=1}^{n}(O_i - P_i)^2}{\sum_{i=1}^{n}(O_i - O_m)^2} \qquad (5)$$

where NSC is Nash-Sutcliffe coefficient, $O_m$ is the mean of the observed values, $O_i$ is the observed values and $P_i$ is the predicted values.

**Results and discussion**

The hourly wind speed data for 3285 days (9 years) were generated using Equation 1 and 2, respectively. The measured cumulative distribution of 78,840 （3285×24） hourly wind speeds was Compared with the distribution generated by Equation 1 and 2 (Figure 2). The overall agreement between measured and simulated cumulative wind speed probability distribution appears satisfactory, with a slight under-prediction in less than 2 m s$^{-1}$ wind speed and a slight over-prediction in 2 to 4.5 m s$^{-1}$ wind speed range by the generation equations. Figure 2 further illustrates that the performance of Equation 1 is little better than that of Equation 2 for reproducing hourly wind speeds. Although the threshold wind speed varies with surface conditions, the threshold is generally assumed as 5 m s$^{-1}$ at the height of 2 m with a erodible field surface condition (Fryrear et al., 1998). For the wind speeds greater than 5 m s$^{-1}$ (erosive wind speed), the agreement appears excellent (Figure 2). The results indicated that the two downscaling models are able to generate hourly wind speeds with high accuracy.

The core module of a wind generator for wind erosion modeling is the cumulative wind speed probability distribution (Figure 2) that is obtained from measured hourly wind speeds (Van Donk et al., 2005). Wind speed data can be generated from the cumulative distribution curve using linear interpolation (Van Donk et al., 2005). The Equation 1 and 2 are capable of reproducing hourly wind speeds magnitude, which further produces the cumulative wind speed probability distribution curve with satisfactory performance (Figure 2), implying that the cumulative distribution curve for generated hourly wind data by Equation 1 or 2 (Figure 2) can also be used in the wind generator for simulating hourly wind speed data when only daily wind statistics data are available for a site.

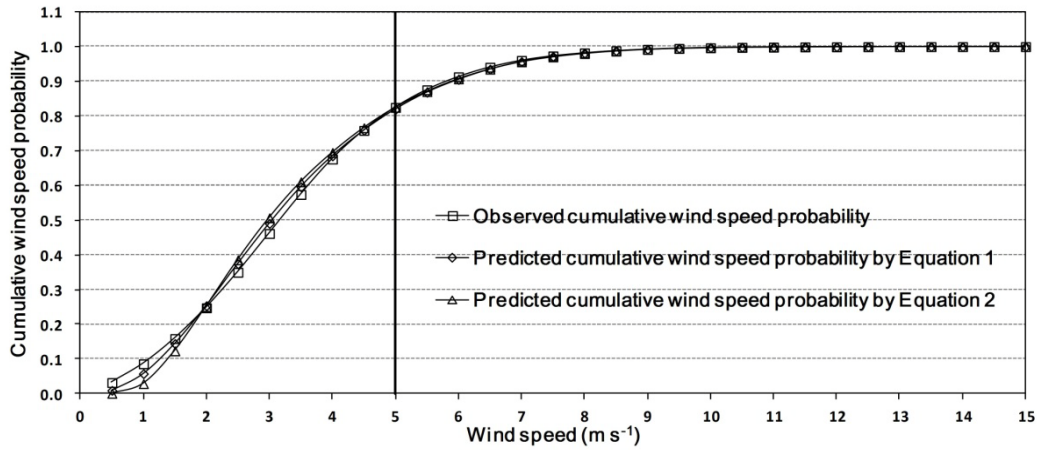

Figure 2. Cumulative distribution of observed hourly wind speed data and predicted hourly wind speed data using Equation 1 and 2 for the 3285 days (9 years).

Daily AWPD values for measured and generated wind speed data were computed using Equations 4. For the 3285-day sampling period, a total of 1489 days had non-zero values of daily AWPD calculated from measured wind data and these days were used to evaluate the performance of the Equation 1 and 2 (Table I ). The average or median for the values of daily AWPD estimated from measured and generated hourly wind speed data change slightly while the maximum varies from 355.44 W m$^{-2}$ to 400.91 W m$^{-2}$ (Table I ).

Table I Statistical characteristic of the daily AWPD values computed from observed and predicted hourly wind speeds for the 1489 days.

| Statistical characteristic | AWPD (average erosive wind power density, W m$^{-2}$) | | |
| --- | --- | --- | --- |
| | Observed | Predicted by Equation 1 | Predicted by Equation 2 |
| Average | 16.46 | 16.78 | 16.96 |
| Median | 5.47 | 5.78 | 5.38 |
| Maximum | 355.44 | 363.02 | 400.91 |
| Minimum | 0.00 | 0.00 | 0.00 |

Comparisons between observed and predicted daily AWPD values are presented in Figure 3. The NSC are 0.92 and 0.81 for Equation 1 and 2, respectively. The predicted daily AWPD fit observed counterparts very well (Figure 3). Good agreement is shown in Figure 5 due to the good performance of the downscaling methods for generating hourly wind speeds (Figure 2). For the two downscaling methods, the Equation 1 requires more inputs (daily average and maximum wind speed), therefore, its performance is better than that of the Equation 2, which only needs daily average wind speed (Figure 3). Overall, the performance of the downscaling methods are also satisfactory for the wind erosivity (AWPD) estimation. The downscaling methods may span the gap between daily wind statistics and hourly wind data. With these methods, we can estimate the wind erosivity from some rough wind data, such as daily wind dataset.

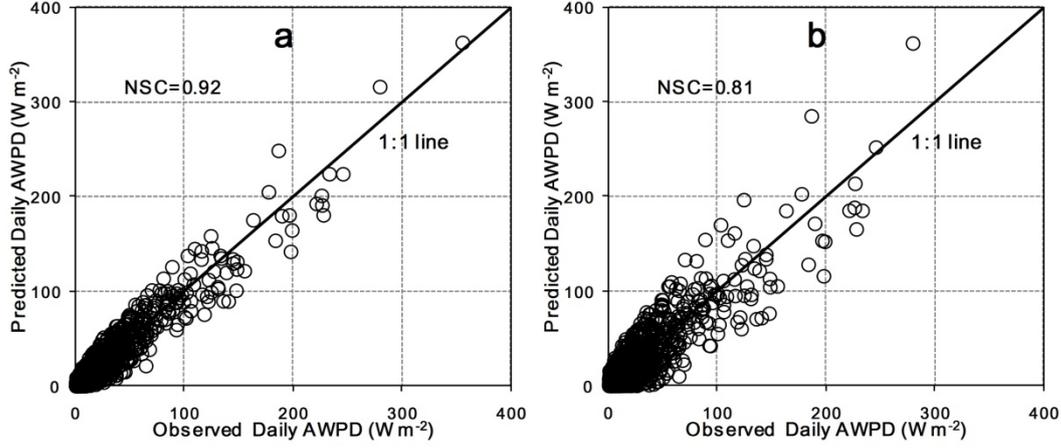

Figure 3. Observed daily AWPD values calculated from measured hourly wind speeds VS (a) predicted daily AWPD values calculated from the generated hourly wind speeds by Equation 1, and (b) predicted daily AWPD values calculated from the generated hourly wind speeds by Equation 2. NSC is the Nash-Sutcliffe coefficient.

The downscaling methods for daily wind statistics can be flexible for the type of inputs. If the hour of the day when wind speed is maximum ($H_{max}$) is also available, we can adjust Equation 1 (Equation 2) as Equation 6 (Equation 7) as:

$$W_n = W_{ave} + \frac{1}{\pi} W_{max} COS(\frac{(n - H_{max}) \cdot \pi}{12}) \qquad (6)$$

$$W_n = W_{ave} + \frac{1}{2} W_{ave} COS(\frac{(n - H_{max}) \cdot \pi}{12}) \qquad (7)$$

Mathematically, for one day, with regard to the magnitude of hourly wind speeds generated by Equation 1 (Equation 2) and Equation 6 (Equation 7), the Equation 1 (Equation 2) is equivalent to Equation 6 (Equation 7). Correspondingly, the cumulative wind speed probability distribution (Figure 2) of the hourly wind speeds generated by Equation 1 (Equation 2) and that generated by Equation 6 (Equation 7) are the same. For some wind erosion models, the time step is generally one day or more than one day for non-single event wind erosion modeling (Hagen, 1996; Fryrear et al, 1998; Webb and McGowan, 2009), the users may be more concerned about the magnitude of hourly wind speeds than the hour when the maximum or minimum wind speed occurs within a day. Thus Equation 1 and 2 can be used to reproduce the magnitude of hourly wind speeds for a day. However, if the users intend to model wind erosion processes for sub-daily step, not only the wind speeds magnitude but also the real diurnal wind pattern is required. In this case, Equation 6 and 7 are more suitable for use in simulating hourly wind speeds curve within a day while the two equations need one more input ($H_{max}$), although the $H_{max}$ for a site may be obtained from meteorological database or publications for the site.

The pattern of diurnal wind speed variation is generally not rigorous cosine (sinusoidal), which may lead a discrepancy between the magnitude of observed and predicted hourly wind speeds for a day (Skidmore and Tatarko, 1990; Ephrath et al., 1996; Donatelli et al., 2009). Therefore, the parameters ($\frac{1}{\pi}$ or $\frac{1}{2}$) of the second term in the Equation 1 (Equation 6) and Equation 2 (Equation 7) may have different values for different geographical regions. Accordingly, it may be risky to use the four formulas to downscale daily wind data for other sites with different wind fluctuations. The results allow researchers to explore new equations based on the formulations of the Equation 1 (Equation 6) and Equation 2 (Equation 7) for downscaling daily wind data. Further studies are needed to validate the downscaling methods using inter-site wind data.

**Conclusion**

In this study, we established a simple method to reproduce hourly wind speed data from daily average and maximum wind speeds together or daily wind speed data only.

3285 days (9 years) hourly wind series of a typical windy region were chosen to evaluate how well wind is generated by Equation 1 and 2. The results presented that the overall agreement between measured and simulated cumulative wind speed probability distribution appears satisfactory, with a slight under-estimation in less than 2 m s$^{-1}$ wind speed and a slight over-estimation in 2 to 4.5 m s$^{-1}$ wind speed range by Equation 1 and 2. The NSC of Equation 1 and 2 are 0.92 and 0.81 for daily AWPD estimation, respectively, further indicating that the downscaling methods are also satisfactory for wind erosivity estimation. It may be risky to directly extend and apply the downscaling methods to the regions with different geographical environment.

**Acknowledges**
The authors would like to thank J. E. Stout and T. M. Zobeck of Wind Erosion and Water Conservation Research Unit, Lubbock, Texas, USA for supplying the wind data used in this paper.